\documentclass[prb,reprint]{revtex4-1}
\pdfoutput=1
\usepackage{graphicx}
\usepackage{bm}
\usepackage{hyperref}
\usepackage{color}

\begin{document}

\title{When are the $q/3$ fractional quantum Hall states stable?}
\date{today}

\author{Yang Liu}
\affiliation{Department of Electrical Engineering,
Princeton University, Princeton, New Jersey 08544}
\author{J.\ Shabani}
\affiliation{Department of Electrical Engineering,
Princeton University, Princeton, New Jersey 08544}
\author{M.\ Shayegan}
\affiliation{Department of Electrical Engineering,
Princeton University, Princeton, New Jersey 08544}

\date{\today}

\begin{abstract}
  Magneto-transport measurements in a wide GaAs quantum well in which
  we can tune the Fermi energy ($E_F$) to lie in different Landau
  levels of the two occupied electric subbands reveal a remarkable
  pattern for the appearance and disappearance of fractional quantum
  Hall states at $\nu$ = 10/3, 11/3, 13/3, 14/3, 16/3, and 17/3. The
  data provide direct evidence that the $q/3$ states are stable and
  strong even at such high fillings as long as $E_F$ lies in a
  ground-state ($N=0$) Landau level of either of the two electric
  subbands, regardless of whether that level belongs to the symmetric
  or the anti-symmetric subband. Evidently, the node in the
  out-of-plane direction of the anti-symmetric subband does not
  de-stabilize the $q/3$ fractional states. On the other hand, when
  $E_F$ lies in an excited ($N>0$) Landau level of either subband, the
  wavefunction node(s) in the in-plane direction weaken or completely
  de-stabilize the $q/3$ fractional quantum Hall states. Our data also
  show that the $q/3$ states remain stable very near the crossing of
  two Landau levels belonging to the two subbands, especially if the
  levels have parallel spins.
\end{abstract}


\maketitle

\section{Introduction}

\begin{figure}
\includegraphics[width=0.45\textwidth]{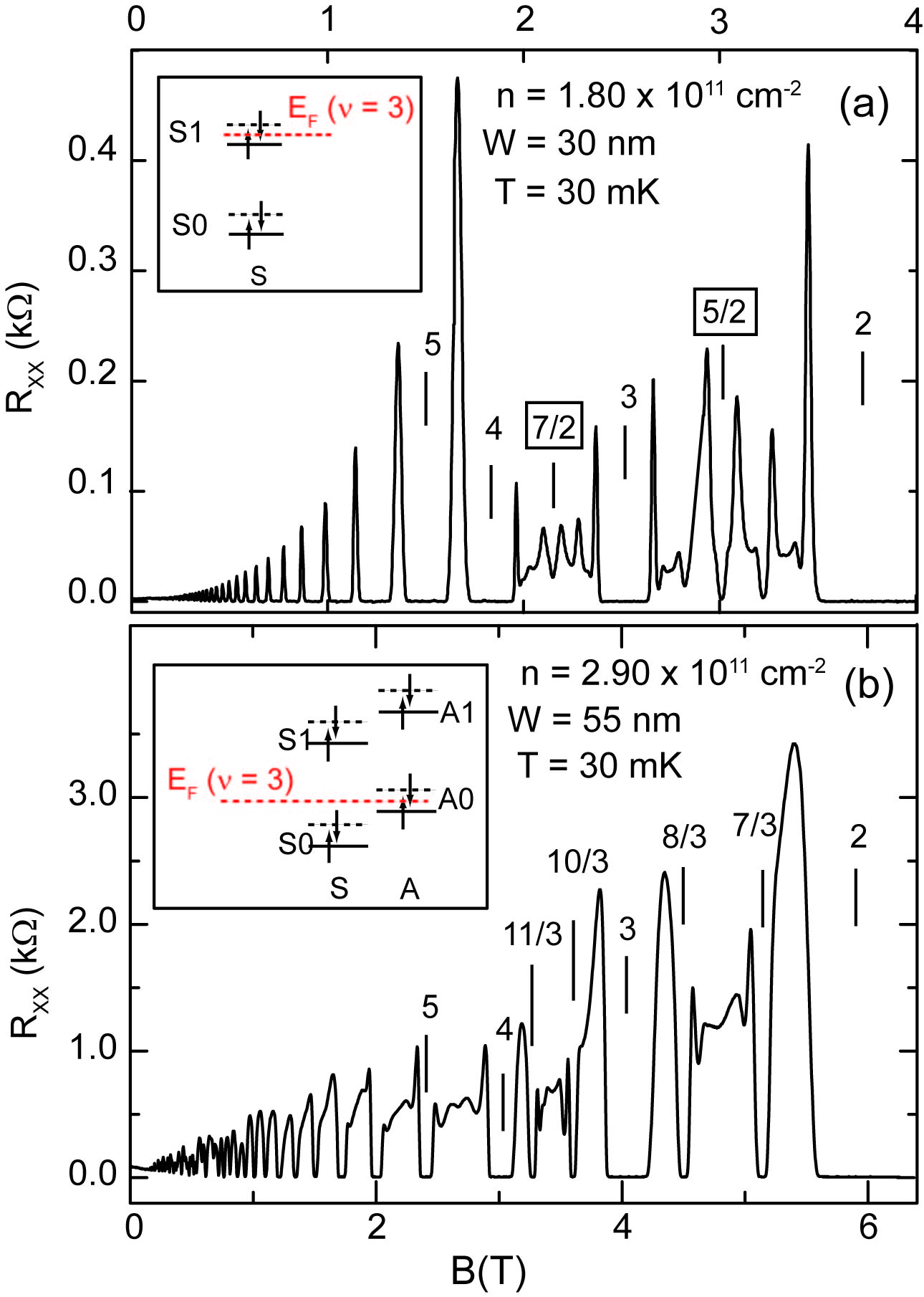}%
\caption[Schematic of electron LLs as a function of increasing charge
imbalance.]{\label{fig:NQWvsWQW} Longitudinal resistance ($R_{xx}$) vs.
  perpendicular magnetic field ($B$) traces are shown for electrons
  confined to: (a) a narrow (well-width $W=$ 30 nm) GaAs quantum well,
  and (b) a wide ($W=$ 55 nm) quantum well. In (a) FQH states at
  $\nu=5/2$ and 7/2 can be clearly seen, but the states at $\nu=7/3$
  and 8/3 are weak. In contrast, the even-denominator states are
  absent in (b) but strong FQH states are seen at $\nu=7/3$ and
  8/3. Note also the absence of FQH states for $\nu > 4$ in (a). The
  insets schematically show the positions of the spin-split LLs of the
  lowest (S) and second (A) electric subbands, as well as the position
  of $E_F$ at $\nu=3$; the indices $N=0$ and $N=1$ indicate the lowest
  and the excited LLs, respectively. The subband separation for the
  trace in (b) is $\Delta=24$ K.}
\end{figure}

The fractional quantum Hall (FQH) effect, \cite{Tsui.PRL.1982}
signaled by the vanishing of the longitudinal resistance and the
quantization of the Hall resistance, is the hallmark of an interacting
two-dimensional electron system (2DES) in a large perpendicular
magnetic field. It is a unique incompressible quantum liquid phase
described by the celebrated Laughlin wavefunction.
\cite{Laughlin.PRL.1983} In a standard, single-subband 2DES confined
to a low-disorder GaAs quantum well, the FQH effect is most
prominently observed at low Landau level (LL) filling factors $\nu<2$,
where the Fermi energy ($E_F$) lies in the spin-resolved LLs with the
lowest orbital index ($N=0$). \cite{CF_Jain} The strongest states are
seen at the $q/3$ fractional fillings, namely at $\nu=1/3$, 2/3, 4/3,
and 5/3. In contrast, as illustrated in Fig. 1(a), when $E_F$ lies in
the second ($N=1$) set of LLs ($2<\nu<4$), the equivalent $q/3$ states
at $\nu = 7/3, 8/3, 10/3,$ and 11/3 are much weaker.
\cite{Pan.PRL.1999,Toke.PRB.2005} In yet higher LLs ($\nu>4$), e.g.,
at $\nu=$13/3, 14/3, 16/3, and 17/3, which correspond to $E_F$ being
in the third ($N=2$) set of LLs, the FQH states are essentially absent;
\cite{Lilly.PRL.1999,Du.SSC.1999,Gervais.PRL.2004} see
Fig. 1(a). This absence is believed to be a result of the larger
extent of the electron wavefunction (in the 2D plane) and its extra
nodes that modify the (exchange-correlation) interaction effects and
favor the stability of various non-uniform charge density states
(e.g., stripe phases) over the FQH states.
\cite{TheQHE,MacDonald.PRB.1986,Ambrumenil.JPC.1988,Koulakov.PRL.1996,Moessner.PRB.1996}

Recently, the FQH effect was examined in a wide GaAs quantum well
where two electric subbands are occupied.\cite{Shabani.PRL.2010} A
main finding of Ref. \onlinecite{Shabani.PRL.2010} is highlighted in
Fig. 1(b): When the Fermi level ($E_F$) lies in the $N=0$ LLs of the
anti-symmetric electric subband, the even-denominator FQH states (at
$\nu=5/2$ and 7/2) are absent and, instead, strong FQH states are
observed at $q/3$ fillings $\nu=$ 7/3, 8/3, 10/3 and 11/3. Here we
extend the measurements in this two-subband system and examine the
stability of the $q/3$ FQH states at even higher fillings as we tune
the position of $E_F$ to lie in different LLs of the two
subbands. At a fixed 2DES density, we observe a remarkable pattern of
alternating appearance and disappearance of the $q/3$ states as we
tune the subband separation and the position of $E_F$. The data
demonstrate that the $q/3$ states are stable even at filling factors
as high as $\nu=17/3$, as long as $E_F$ lies in a ground state ($N=0$)
LL, regardless of whether that LL belongs to the symmetric or
anti-symmetric subband.


\section{Sample and experimental details}

\begin{figure*}
\includegraphics[width=.9\textwidth]{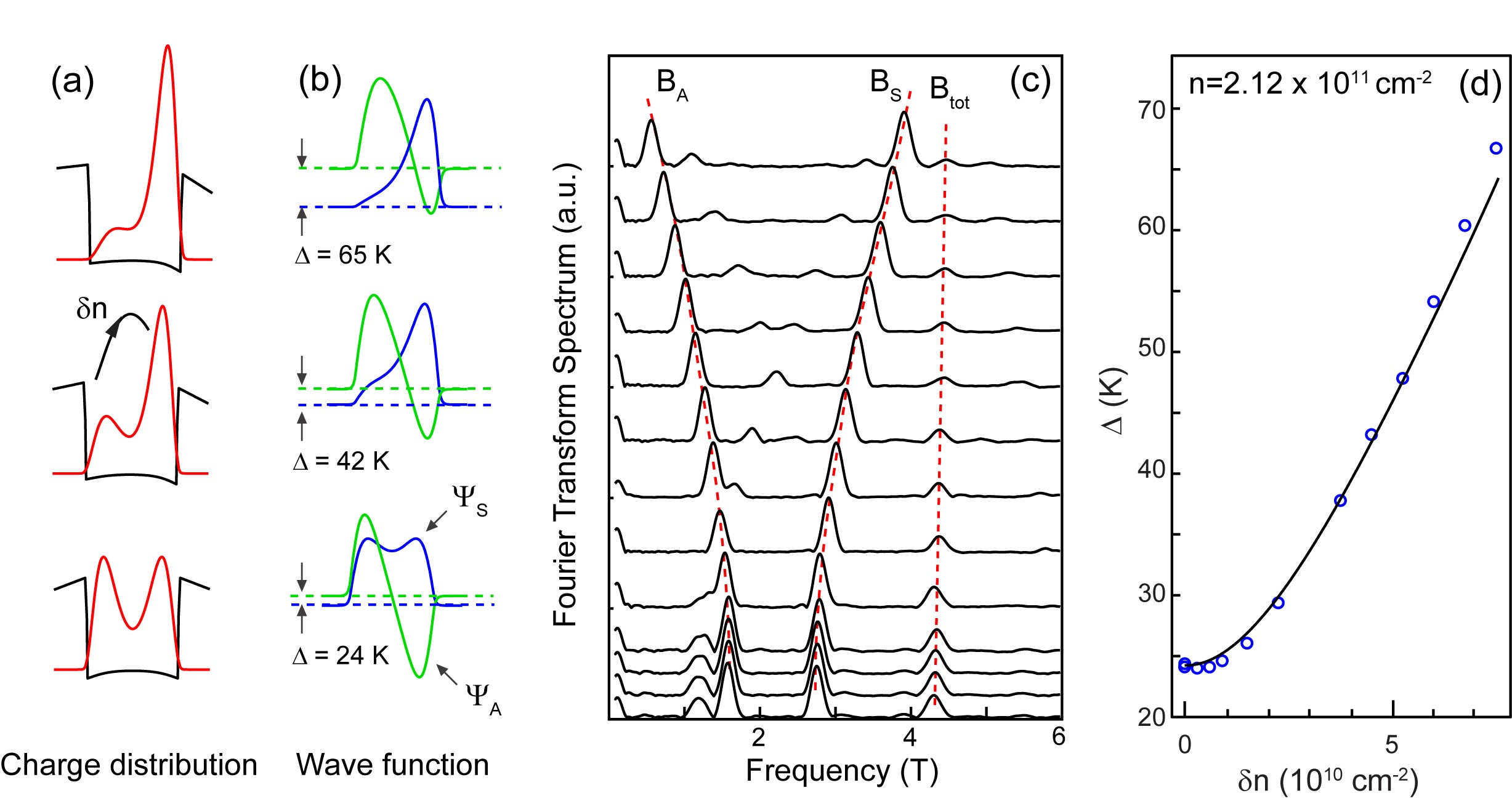}%
\caption[Schematic of electron LLs as a function of increasing charge
imbalance.]{\label{fig:cartoon} (a) Charge distribution (red) and
  potential (black), and (b) wave functions from self-consistent
  simulations for a 55 nm-wide GaAs QW. The charge density is kept
  fixed at $n=2.12\times 10^{11}$cm$^{-2}$. The subband separation
  $\Delta$ is the smallest when the QW is balanced (bottom panels),
  and increases as the QW is imbalanced. (c) The Fourier transform
  spectra of the measured low-field Shubnikov-de Haas
  oscillations. Each spectrum exhibits two main peaks, denoted as $B_A$ and
  $B_S$, whose separation increases as the QW is imbalanced (from
  bottom to top). (d) The subband separation $\Delta$ determined from
  the Fourier transforms through $\Delta=\frac{\hbar
    e}{m^{\star}}(B_S-B_A)$, plotted as a function of the charge
  distribution asymmetry $\delta n$. The solid curve represents
  $\Delta$ vs. $\delta n$ from self-consistent calculations for a 55
  nm-wide GaAs QW.}
\end{figure*}

Our sample, grown by molecular beam epitaxy, is a 55 nm-wide GaAs
quantum well (QW) bounded on each side by undoped
Al$_{0.24}$Ga$_{0.76}$As spacer layers and Si $\delta$-doped
layers.\footnote{Our sample is the same as the one used in
  Ref. \onlinecite{Shabani.PRL.2010}.  Based on our careful
  measurements of the subband separation ($\Delta$) while imbalancing
  the QW (see Fig.~\ref{fig:cartoon}), we conclude that the QW has a
  width of 55 nm, slightly smaller than 56 nm which was quoted in
  Ref. \onlinecite{Shabani.PRL.2010}. We emphasize that throughout our
  paper we use the experimentally measured values of $\Delta$, and
  that the exact width of the QW has no bearing on our
  conclusions.} We fitted the sample with an evaporated Ti/Au
front-gate and an In back-gate to change the 2D electron density, $n$,
and tune the charge distribution symmetry and the occupancy of the two
electric subbands, as demonstrated in Fig.~\ref{fig:cartoon}. This
tunability, combined with the very high mobility ($\sim$ 400 m$^2$/Vs)
of the sample, is key to our success in probing the strength of the
$q/3$ states at high fillings.

\begin{figure}
\includegraphics[width=0.45\textwidth]{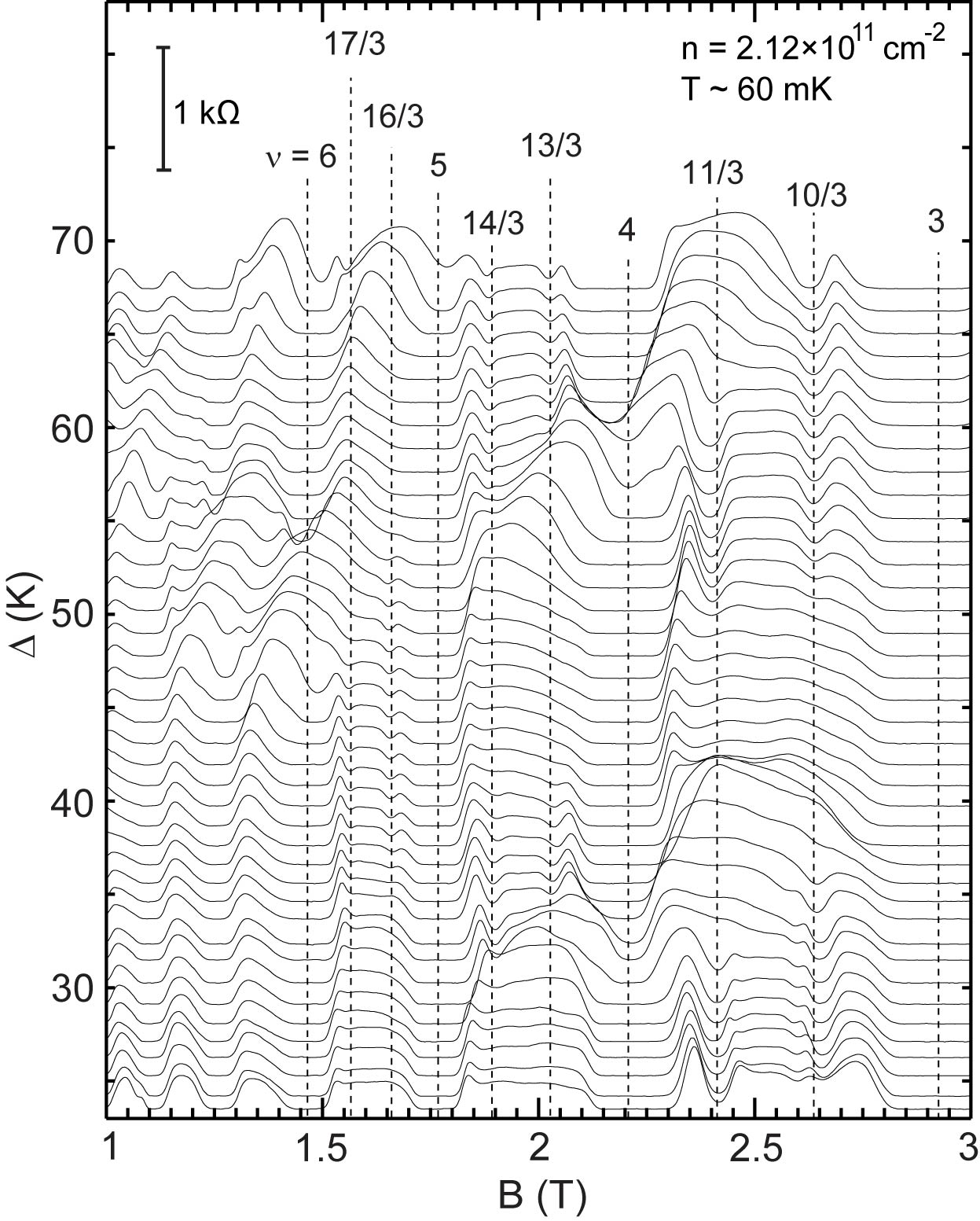}
\caption{\label{fig:waterfall} Waterfall plot of $R_{xx}$
    vs. $B$ taken at a fixed density $n=2.12\times
      10^{11}$cm$^{-2}$ as the subband separation ($\Delta$) is
    increased. The scale for $R_{xx}$ is indicted in the upper left
  (0 to 1 k$\Omega$). Each trace is shifted vertically so that its
  zero (of $R_{xx}$) is aligned with its measured value of $\Delta$
  which is used as the y-axis of the waterfall plot. Vertical lines
  mark the field positions of the filling factors, $\nu$. }
\end{figure}

When the QW in our experiments is ''balanced'', i.e., the charge
distribution is symmetric, the occupied subbands are the symmetric (S)
and anti-symmetric (A) states (see the lower panels in Figs. 2(a) and
(b)). When the QW is ''imbalanced,'' the two occupied subbands are no
longer symmetric or anti-symmetric; nevertheless, for brevity, we
still refer to these as S (ground state) and A (excited state). In our
experiments, we carefully control the electron density and charge
distribution symmetry in the QW via applying back- and front-gate
biases.\cite{Suen.PRL.1994,Shabani.PRL.2009} For each pair of gate
biases, we measure the occupied subband electron densities from the
Fourier transforms of the low-field ($B\le 0.5$ T) Shubnikov-de Haas
oscillations. These Fourier transforms, examples of which are shown in
Fig. 2(c), exhibit two peaks ($B_S$ and $B_A$) whose frequencies,
multiplied by $2e/h$, give the subband densities, $n_S$ and $n_A$. The
difference between these densities directly gives the subband
separation, $\Delta$, through the expression
$\Delta=\frac{\pi\hbar^2}{m^{*}}(n_S-n_A)$, where $m^{*}$ is the
electron effective mass. Note that, at a fixed total density, $\Delta$
is smallest when the charge distribution is balanced and it increases
as the QW is imbalanced. Figure 2(d) shows the measured $\Delta$ as a
function of the charge $\delta n$ transferred between the back and
front sides of the QW. Note that we measure $\delta n$ from the change
in the sample density induced by the application of either the
back-gate or the front-gate bias.


\section{magneto-transport data}

\begin{figure*}
\includegraphics[width=.9\textwidth]{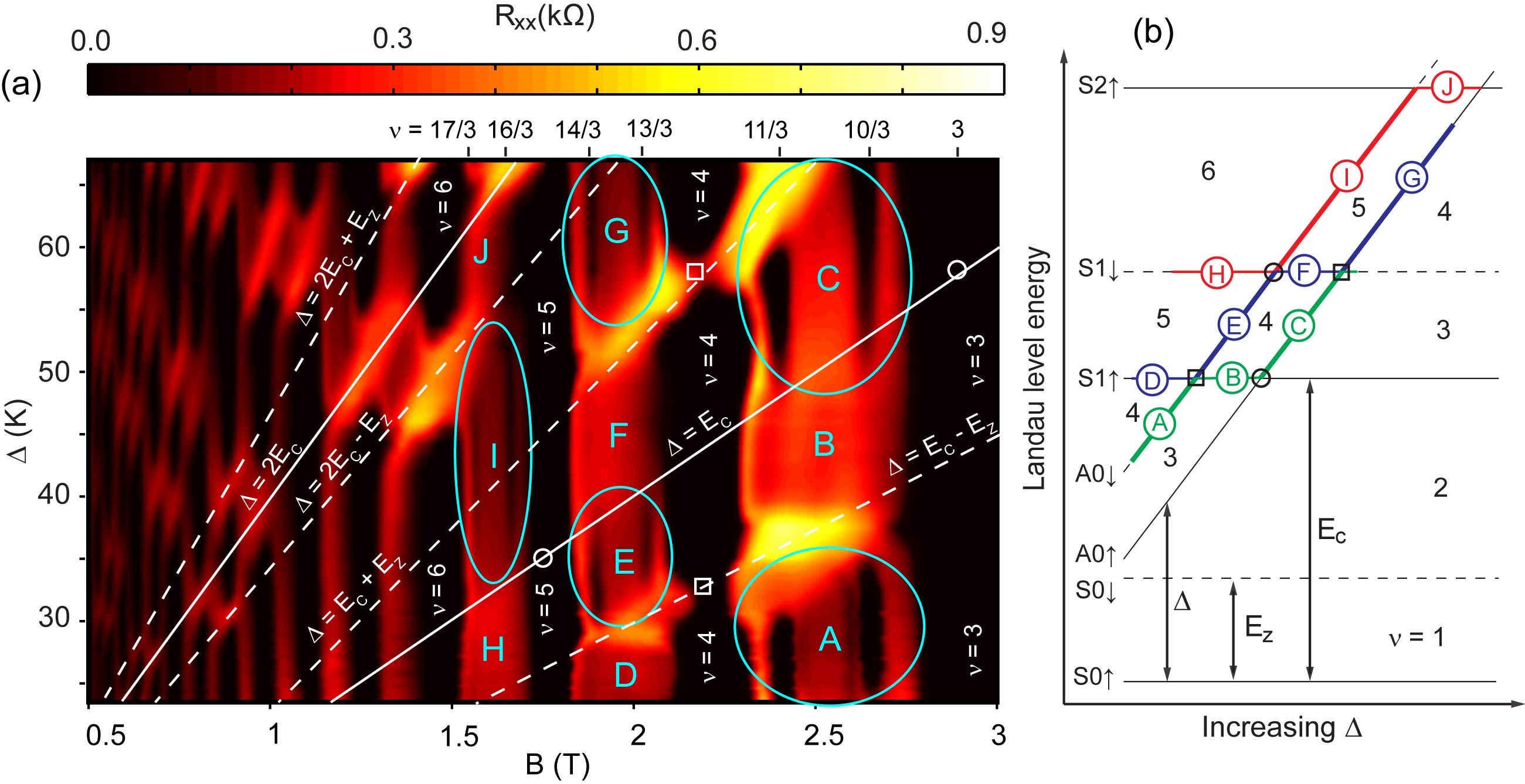}%
\caption{\label{fig:colorful} Evolution of $R_{xx}$ data taken at a
  fixed density $n=2.12\times 10^{11}$cm$^{-2}$ as the subband
  separation ($\Delta$) is increased. (a) A color-scale plot of the
  data shown in Fig.~\ref{fig:waterfall}. The dark regions are where
  the integer or fractional quantum Hall states are observed at the
  indicated values of $\nu$. The solid white lines denote $\Delta=E_C$
  and $\Delta=2 E_C$, where $E_C$ is the cyclotron energy. The dashed
  white lines are drawn such that they pass through the even-filling
  coincidences (see text). The cyan ellipses mark different regions
  (A, C, E, G, and I) where FQH states are seen. (b) Schematic
  electron Landau level diagram as a function of increasing
  $\Delta$. The relevant energies, $\Delta$, the cyclotron energy
  ($E_C$), and the Zeeman energy ($E_Z$), are shown. The position of
  the Fermi level is plotted in different colors for several filling
  factor regions: 3 $< \nu <$ 4 (green), 4 $< \nu <$ 5 (blue), and 5
  $< \nu <$ 6 (red). The letters correspond to the regions in the (a)
  panel; the regions where FQH states are observed are marked by
  thicker lines.}
\end{figure*}

\begin{figure}
\includegraphics[width=0.4\textwidth]{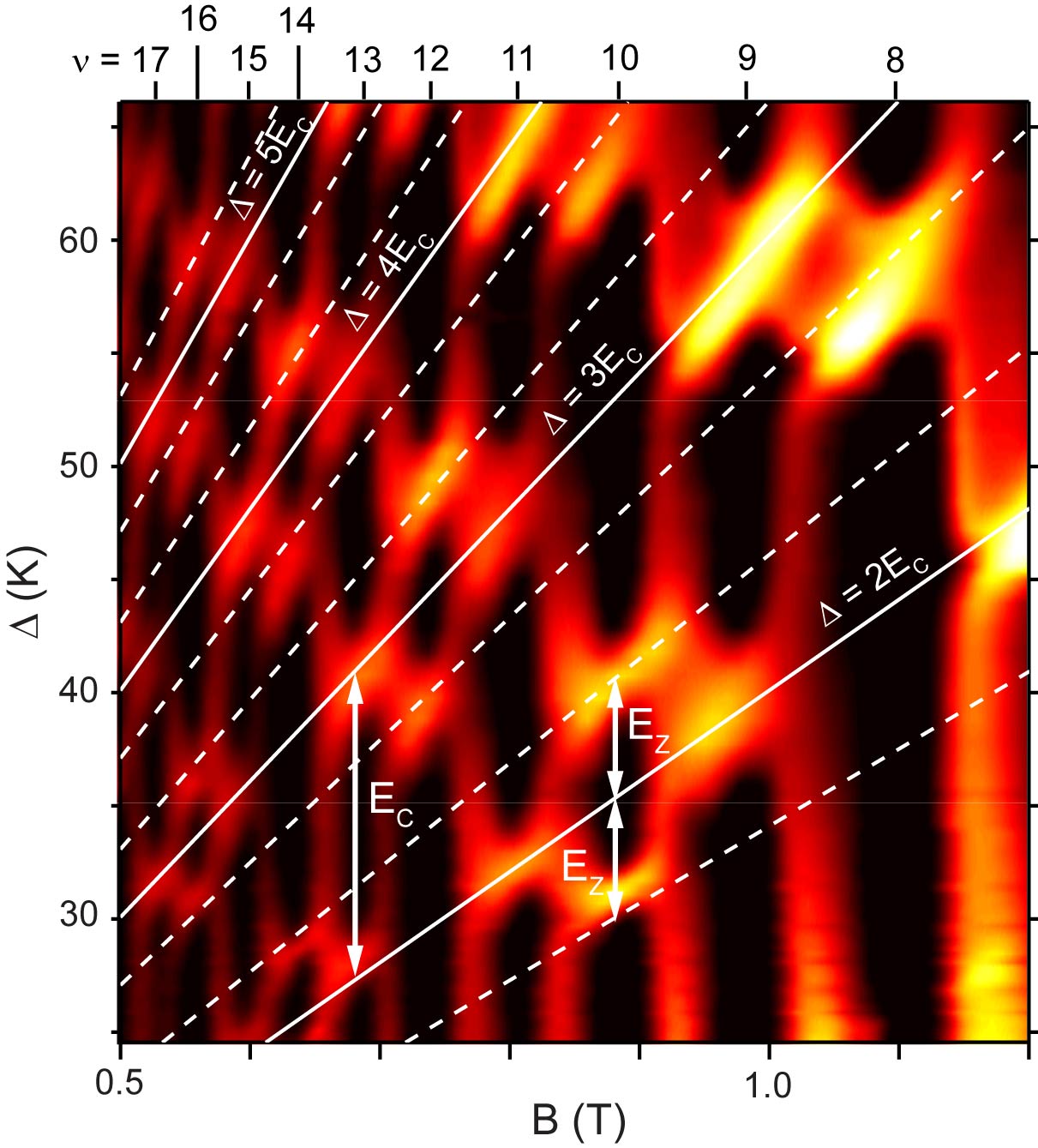}
\caption{\label{fig:low_field_colorful} An expanded color-scale plot
  of $R_{xx}$ data at low fields for $n=2.12\times 10^{11} $
  cm$^{-2}$. The solid white lines denote $\Delta=iE_C$ for $i=2,3,4,
  5$. The dashed white lines represent $\Delta=iE_C\pm E_Z$, using a
  fixed $g^{\star}=8.8$ (see text).}
\end{figure}

\begin{figure*}
\includegraphics[width=.9\textwidth]{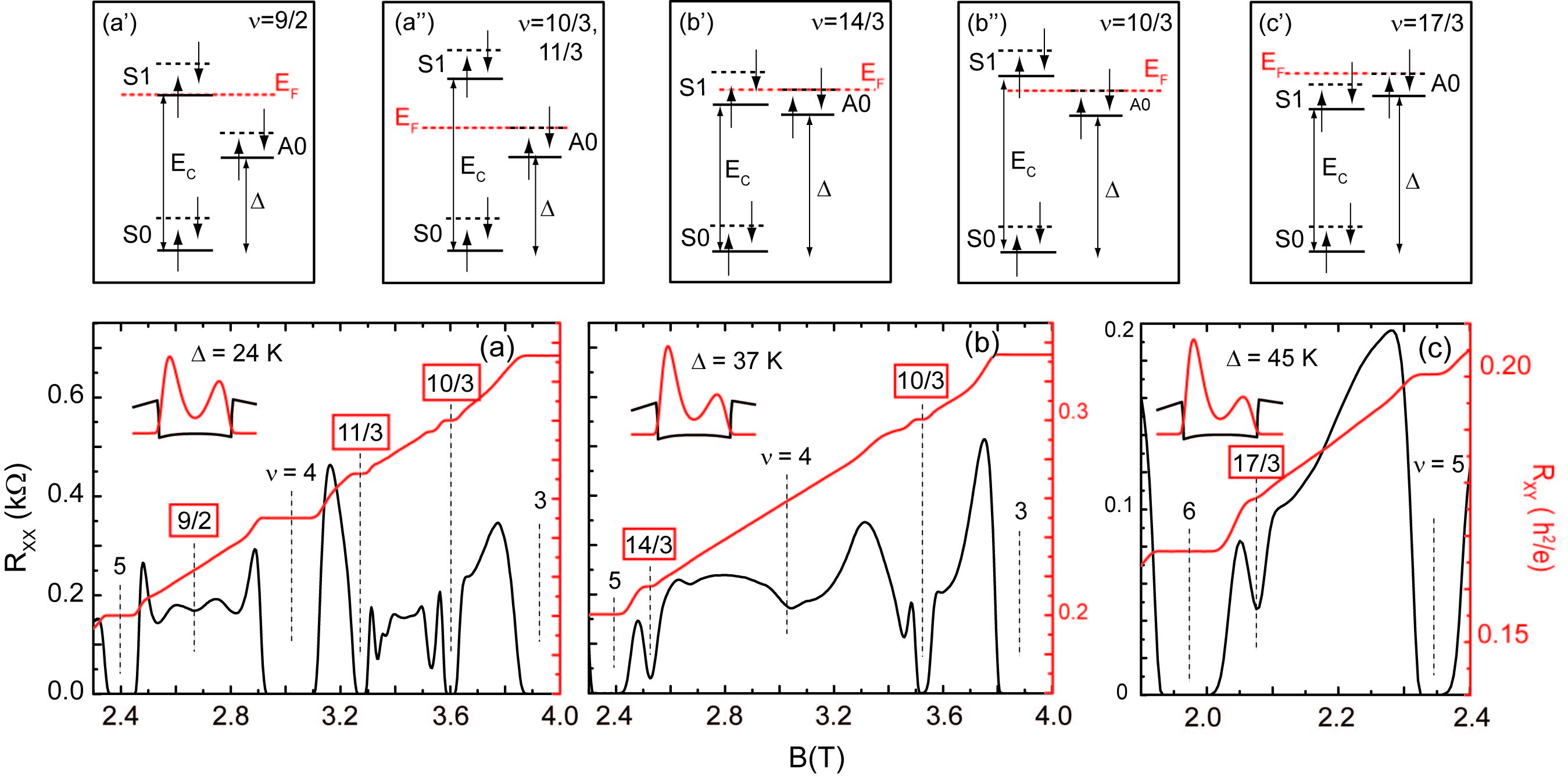}%
\caption[Schematic of electron LLs as a function of increasing charge
imbalance.]{\label{fig:hall_trace_290} (a), (b), (c) Longitudinal
  ($R_{xx}$) and Hall ($R_{xy}$) resistances for the 55 nm-wide QW at
  a higher density $n=2.90\times 10^{11} $ cm$^{-2}$. Panels (a) and
  (b) share the same scales for $R_{xx}$ and $R_{xy}$. The traces were
  taken at $T=30$ mK, and the insets show the charge distributions
  calculated at $B=0$. The LL diagrams for fractional fillings in
  panels (a), (b) and (c) are shown in (a') and (a''), (b') and (b''),
  and (c'), respectively. Fractional quantum Hall states at
  $\nu=10/3$, 11/3, 14/3, and 17/3 are clearly seen when $E_F$ lies in
  either the A0$\uparrow$ or A0$\downarrow$ levels.}
\end{figure*}

\begin{figure}
\includegraphics[width=0.45\textwidth]{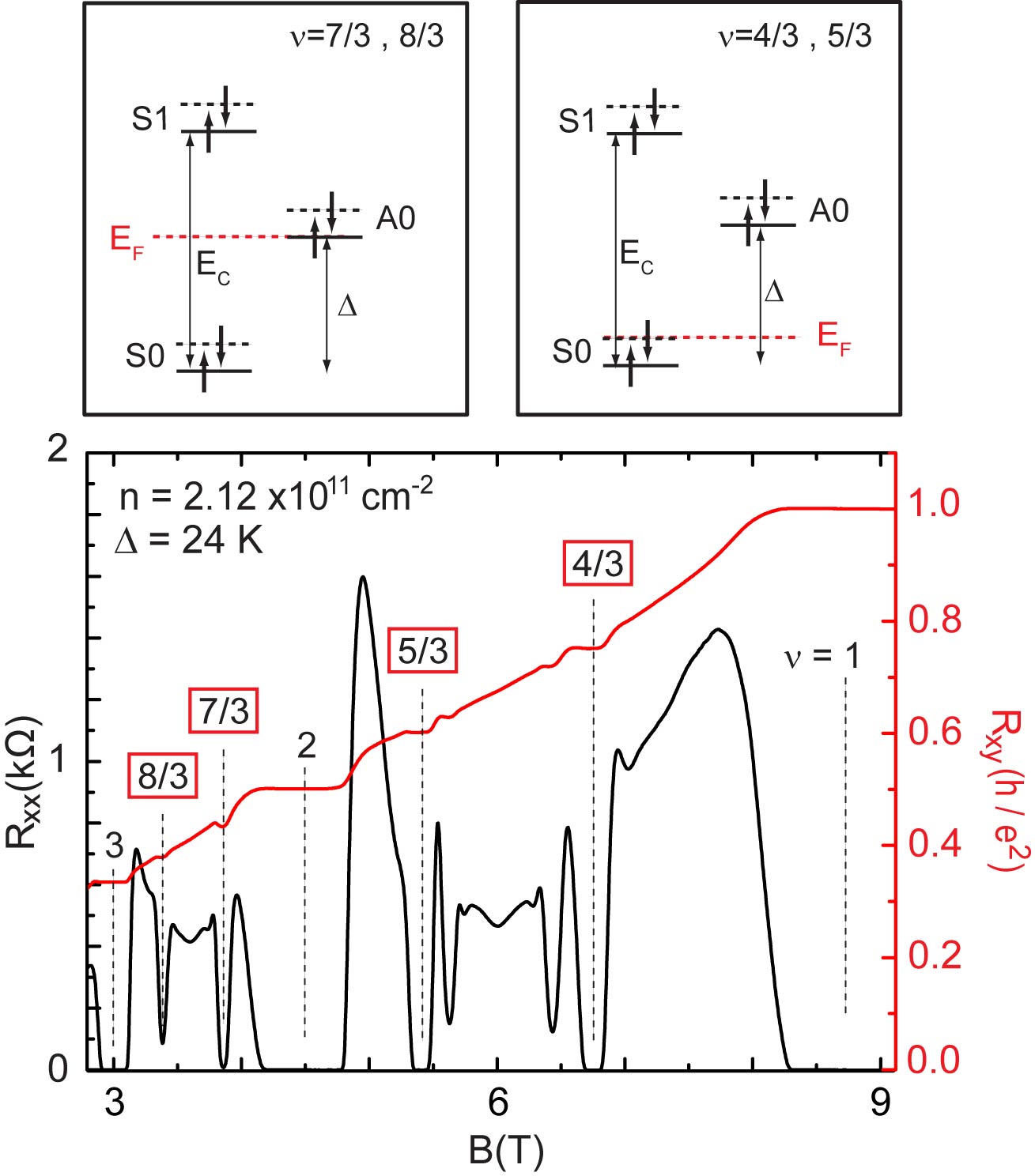}%
\caption[Schematic of electron LLs as a function of increasing charge
imbalance.]{\label{fig:hall_trace_212} $R_{xx}$ and
  $R_{xy}$ traces at high magnetic fields and $T=30$ mK
  for $n=2.12\times 10^{11} $ cm$^{-2}$ for the "balanced"
  QW. Fractional quantum Hall states at $\nu=4/3$, 5/3, 7/3, and 8/3
  are clearly seen. The upper panels show the LL diagrams and
  positions of $E_F$ for the indicated fillings.}
\end{figure}
 
Figure~\ref{fig:waterfall} shows a series of longitudinal resistance
($R_{xx}$) vs.~magnetic field ($B$) traces taken at a fixed density
$n=2.12\times 10^{11} $ cm$^{-2}$ as the subband spacing is
increased. The y-axis is $\Delta$, which is measured from the
low-field Shubnikov-de Haas oscillations of each trace. The same data
are interpolated and presented in a color-scale plot in
Fig.~\ref{fig:colorful}(a). In Fig.~\ref{fig:low_field_colorful}, we
show a color-scale plot of the data in the low field regime.

In Figs.~\ref{fig:waterfall}, \ref{fig:colorful}(a), and
\ref{fig:low_field_colorful} we observe numerous LL coincidences at
various integer filling factors, signaled by a weakening or
disappearance of the $R_{xx}$ minimum.  For example, the $R_{xx}$
minimum at $\nu=4$ is strong and wide at all values of $\Delta$ except
near $\Delta$ = 32 and 58 K, marked by squares in
Fig.~\ref{fig:colorful}(a), where it becomes narrow or
disappears. Such coincidences can be easily explained in a simple fan
diagram of the LL energies in our system as a function of increasing
$\Delta$, as schematically shown in Fig.~\ref{fig:colorful}(b). In
this figure, we denote an energy level by its subband index (S or A),
LL index ($N=0, 1, 2, \cdots$), and spin ($\uparrow$ or
$\downarrow$). Also indicated in Fig.~\ref{fig:colorful}(b) are the
separations between various levels: the cyclotron energy ($E_C=\hbar
eB/m^{*}$), Zeeman energy ($E_Z=g^{*}\mu_BB$, where $g^{*}$ is the
effective Land\'e g-factor), and $\Delta$. From
Fig.~\ref{fig:colorful}(b) it is clear that the condition for observing
a LL coincidence at odd fillings is $\Delta=iE_C$, while for
coincidences at even fillings, the condition is $\Delta=iE_C\pm E_Z$;
in both cases, $i$ is a positive integer.

In Figs.~\ref{fig:colorful}(a) and \ref{fig:colorful}(b), we have
indicated the two coincidences at $\nu=4$ with squares. Note that the
coincidences at even fillings correspond to a crossing of two levels
with antiparallel spins. In Figs.~\ref{fig:waterfall} and
\ref{fig:colorful}(a), the coincidences at low, odd fillings (e.g.,
$\nu=3$ and 5) are not as easy to see at low temperatures since the
resistance minima remain strong as the two LLs, which have parallel
spins, cross. Such behavior has been reported previously and has been
interpreted as a signature of easy-plane ferromagnetism.
\cite{Jungwirth.PRL.1998,Muraki.PRL.2001,Vakili.PRL.2006} We note that
our data taken at higher temperatures ($T$ = 0.31 K) reveal a
weakening of the $\nu=5$ minimum at $\Delta=35$ K, and of the $\nu=3$
minimum at $\Delta=58$ K;\cite{Shabani.Thesis.2011} these are marked
by circles in Fig.~\ref{fig:colorful}(a). The crossings at higher odd
fillings are clearly seen in Figs. ~\ref{fig:colorful}(a) and
\ref{fig:low_field_colorful}; e.g., the $\nu=7$ minimum disappears at
around $\Delta=50$ K, and $\nu=9$ around $\Delta=$ 40 K and 60
K.\footnote{We note that when the charge distribution is nearly
  symmetric, LL coincidences at even fillings are also difficult to see at
  very low temperatures. For example, there is a coincidence at
  $\nu=8$ at $\Delta\simeq 26$ K but we can only see a weakening of the
  $R_{xx}$ minimum at $T\gtrsim 0.3$ K.}

In Figs.~\ref{fig:colorful}(a) and \ref{fig:low_field_colorful} we
include several solid white lines representing $\Delta=iE_C$, assuming
GaAs band effective mass of $m^{*}=0.067$ (in units of free electron
mass). These lines indeed pass through the positions of the $observed$
LL coincidences for odd fillings, implying that $\Delta$ is not
re-normalized at LL coincidences. We note that, with
the application of magnetic field, the subband electron occupation
might vary because of the finite number of discrete LLs that are
occupied. This could lead to a redistribution of charge which in
turn could lead to changes in $\Delta$ as a function of magnetic
field. At LL coincidences, however, the two crossing LLs which
belong to the different subbands are energetically degenerate. If
the coincidence occurs at the Fermi energy, electrons can move
between the two degenerate LLs so that the subband occupancy and the
charge distribution, and therefore $\Delta$, are restored back to
their zero-field values. This conjecture is indeed confirmed by
self-consistent calculations reported for a two-subband 2D electron
system in a perpendicular magnetic field:\cite{Trott.PRB.1989}
While the subband occupancy and $\Delta$ oscillate with field, they
equal their zero-field values whenever two LLs belonging to
different subbands coincide at $E_F$. We conclude that the field
positions of the LL coincidences at $E_F$ are determined by the
value of $\Delta$ at $B=0$, and that the lines drawn in Figs. 4(a)
and 5 accurately describe the positions of these coincidences.

The dashed lines in Figs.~\ref{fig:colorful}(a) and
\ref{fig:low_field_colorful}, represent $\Delta=iE_C\pm E_Z$, $i=1,2,
...$, where $g^{*}$ is chosen as a fitting parameter so that these
lines pass through the even-filling coincidences. All the dashed lines
in Figs.~\ref{fig:colorful}(a) and \ref{fig:low_field_colorful} are
drawn using $g^{*}=$ 8.8, except for the $\Delta = E_C \pm E_Z$ lines,
which are drawn using $g^{*}=$ 8.9 and 7.6, respectively. We conclude
that $g^{*}$ is enhanced by a factor of $\sim$ 20 relative to the GaAs
band g-factor (0.44). This enhancement is somewhat larger than the
values reported for GaAs QWs with two subbands occupied. For example,
Muraki $et\ al.$ \cite{Muraki.PRL.2001} reported a $\sim$ 10-fold
enhancement of $g^{*}$ for electrons in a 40 nm-wide QW with $n\sim
3\times 10^{11} $ cm$^{-2}$ while Zhang $et\ al.$
\cite{Zhang.PRB.2006} measured a $\sim$ 5-fold enhancement in a 24
nm-wide QW with $n\sim 7\times 10^{11} $ cm$^{-2}$. It appears then
that the enhancement depends on the QW width and electron density, and
a systematic study of the enhancement would be an interesting future
project. However, we would like to emphasize that the dashed lines in
Figs. \ref{fig:colorful}(a) and \ref{fig:low_field_colorful} pass
through nearly all of the observed coincidences quite well. Since each of
these lines are drown using very similar $g^{*}$, the data imply that
the enhancement is nearly independent of the filling
factor.\footnote{The observation of a significantly enhanced g-factor
  which is independent of the filling factor has been reproted in the
  past [S. J. Papadakis, E. P. De Poortere, and M. Shayegan,
  Phys. Rev. B \textbf{59}, R12743 (1999); Y. P. Shkolnikov, E. P. De
  Poortere, E. Tutuc, and M. Shayegan, Phys. Rev. Lett. \textbf{89},
  226805 (2002)].}

We now focus on the main finding of our work, namely the
correspondence between the stability of the FQH states and the
position of $E_F$. Note in Figs.~\ref{fig:waterfall} and
\ref{fig:colorful}(a) that FQH states are observed only in certain
ranges of $\Delta$. For example, the $\nu=10/3$ and 11/3 states are
seen in the regions marked by A and C in Fig.~\ref{fig:colorful}(a) but
they are essentially absent in the B region. The $\nu=13/3$ and 14/3
states, on the other hand, are absent in regions D and F while they
are clearly seen in regions E and G.

To understand this behavior, in the fan diagram of
Fig.~\ref{fig:colorful}(b) we have highlighted the position of $E_F$ as
a function of $\Delta$ for different filling factors by color-coded
lines.  Concentrating on the range $3<\nu<4$ (green line in
Fig.~\ref{fig:colorful}(b)), at small values of $\Delta$ (region A),
$E_F$ lies in the A0$\downarrow$ level. At higher $\Delta$, past the
first $\nu=4$ coincidence which occurs when $\Delta=E_C - E_Z$, $E_F$
is in the S1$\uparrow$ level (region B). Once $\Delta$ exceeds $E_C$,
$E_F$ lies in the A0$\uparrow$ level (region C) until the second
$\nu=4$ coincidence occurs when $\Delta=E_C + E_Z$. Note in
Fig.~\ref{fig:colorful}(a) that strong FQH states at $\nu=10/3$ and
$11/3$ are seen in regions A and C. From the fan diagram of Fig.
\ref{fig:colorful}(b) it is clear that in these regions $E_F$ is in the
$ground$-$state$ ($N=0$) LLs of the asymmetric subband, i.e.,
A0$\uparrow$ and A0$\downarrow$. In contrast, in region B, where the
10/3 and 11/3 states are essentially absent, $E_F$ lies in an
$excited$ ($N=1$) LL, namely, S1$\uparrow$. We conclude that the 10/3
and 11/3 FQH states are stable and strong when $E_F$ lies in a ground-state LL.

The data in the range $4<\nu<5$ corroborate the above conclusion. In
Fig.~\ref{fig:colorful}(b) we represent the position of $E_F$ in this
filling range by a blue line. In regions E and G, $E_F$ lies in the
ground-state LLs of the asymmetric subband (A0$\downarrow$ and
A0$\uparrow$), and these regions are indeed where the $\nu=13/3$ and
14/3 FQH states are seen. In regions D and F, on the other hand, $E_F$
is in the excited LLs of the symmetric subband (S1$\uparrow$ and
S1$\downarrow$), and the 13/3 and 14/3 FQH states are absent. Data at
yet higher fillings ($5<\nu<6$) follow the same trend: FQH states at
$\nu=16/3$ and 17/3 are seen in region I when $E_F$ is in the
A0$\downarrow$ level,\footnote{\label{footnote1}The $E_C+E_Z$ line
  going through region I does not correspond to a LL coincidence at
  the Fermi energy in this region; this should be evident from
  Fig.~\ref{fig:colorful}(b) diagram. The same is true about the
  $2E_C-E_Z$ line as it goes through region G.} but they are absent in
regions H or J where $E_F$ lies in the S1$\downarrow$ or S2$\uparrow$
levels.

In Fig.~\ref{fig:hall_trace_290} we show additional data for a density
of $n=2.90\times 10^{11} $ cm$^{-2}$ in the same QW. Longitudinal and
Hall resistance traces are shown in the bottom panels for three different
values of $\Delta$, and in each panel the calculated charge
distribution (at $B=0$) is also shown. In the top panels, we show the positions
of the LLs and $E_F$, corresponding to the filling factors in the bottom
panels. In all cases, strong $q/3$ FQH states are observed when $E_F$
lies in the $N=0$ of the A0$\downarrow$ level. Note that the data
shown in Fig. \ref{fig:hall_trace_290} are for asymmetric charge
distributions. We would like to emphasize that strong $q/3$ states are
also observed for symmetric ("balanced") charge distributions; e.g.,
see the bottom trace in Fig. 3, or the traces in
Fig. 2(c) of Shabani $et$ $al.$ \cite{Shabani.PRL.2010}

Next we address the FQH states observed at lower $\nu$ ($< 3$) in our
sample. Data are shown for $n=2.12\times 10^{11} $ cm$^{-2}$ for the
"balanced" QW ($\Delta=23$ K) in Fig.~\ref{fig:hall_trace_212}; the
$R_{xx}$ trace is an extension of the lowest trace shown in
Fig.~\ref{fig:waterfall}. In the range 1 $<\nu<$ 3, strong FQH states
are seen at $\nu=$ 4/3, 5/3, 7/3 and 8/3.  Data taken at yet higher
magnetic fields (not shown) reveal the presence of a very strong FQH
state at $\nu$ = 2/3. From the fan diagram of
Fig.~\ref{fig:colorful}(b), it is clear that $E_F$ at these fillings
lies in an $N=0$ LL, namely, the A0$\uparrow$ ($\nu=$ 7/3 and 8/3),
S0$\downarrow$ ($\nu=$ 4/3 and 5/3), or S0$\uparrow$ ($\nu=$ 2/3)
levels.\footnote{Traces taken at higher values of $\Delta$ reveal that
  the $\nu=$ 7/3 and 8/3 states remain strong up to the $\nu=$ 3
  coincidence. Past this coincidence, the 7/3 and 8/3 states become
  weaker, consistent with the fact that $E_F$ now lies in an excited
  LL (the S1 level, see Fig.~\ref{fig:colorful}(b)).}

\section{discussion}

Our observations provide direct evidence that the $q/3$ FQH states are
strong when $E_F$ resides in a ground-state ($N=0$) LL, regardless of
whether that LL belongs to the A or S subband. This finding implies
that the node in the wavefunction in the $out$-$of$-$plane$ direction
does not significantly de-stabilize the $q/3$ FQH states. On the other
hand, when $E_F$ lies in an $N>0$ LL, the wavefunction node(s) in the
$in$-$plane$ direction weaken or completely de-stabilize the $q/3$ FQH
states. These conclusions are consistent with the data from
single-subband samples,
\cite{Pan.PRL.1999,Lilly.PRL.1999,Du.SSC.1999,Gervais.PRL.2004} as
well as theoretical calculations.  \cite{TheQHE,MacDonald.PRB.1986,
  Ambrumenil.JPC.1988, Koulakov.PRL.1996, Moessner.PRB.1996,
  Toke.PRB.2005} In a composite Fermion picture, our data also imply
that the lower lying (fully occupied) LLs are essentially inert and
the composite Fermions are formed in the partially filled LL where
$E_F$ lies. The composite Fermions, however, could have a spin and/or
subband degree of freedom, as we briefly discuss in the last paragraph
of this section (see also, Ref. \onlinecite{Shabani.PRL.2010}).

Our data also allow us to assess the stability of the FQH states as
two LLs approach each other. In Fig.~\ref{fig:colorful}(a) the dashed
line denoted $E_C-E_Z$ marks the position of the expected crossing
between the A0$\downarrow$ and the S1$\uparrow$ levels, based on the
LL coincidence we observe for the $\nu=4$ quantum Hall state. It is
clear in Fig.~\ref{fig:colorful}(a) that as we approach this line from
the A region, the 10/3 and 11/3 FQH states disappear when $\Delta$ is
about 5 K away from $E_C-E_Z$. A similar statement can be made
regarding the stability of the 11/3 state as the $E_C+E_Z$ dashed line
is approached from the C region, and the stability of the 13/3 and
14/3 states as one approaches the $E_C+E_Z$ line from the G region or
the $E_C-E_Z$ line from the E region.\cite{Note4} Note that what
is common to all these observations is that the boundaries marked by
the dashed lines correspond to the crossing of two LLs with
$antiparallel$ spins.

Data of Fig. \ref{fig:colorful}(a) suggest that, when the two
approaching LLs have $parallel$ spins, the $q/3$ states remain stable
even closer to the expected LL crossings. For example, the 10/3 and
11/3 FQH states in region C are stable very close to the boundary (the
line marked $E_C$) separating this region from B. Similarly, the 13/3
and 14/3 states are stable in region E close to the $E_C$ line
separating E from F. Note that in both cases, i.e., traversing from C
to B or from E to F, the two approaching LLs have parallel spins (see
Fig. \ref{fig:colorful}(b)). We conclude that the relative spins of
the two approaching LLs also play a role in the stability of the $q/3$
FQH states. It is worth emphasizing that, as is evident from
Figs.~\ref{fig:waterfall} and \ref{fig:colorful}(a) data, the relative
spins of the two approaching LLs also play a crucial role in the
stability of the $integer$ quantum Hall (IQH) states. For
antiparallel-spin LLs, the IQH state (e.g., at $\nu=4$) becomes very
weak or completely disappears, while for the parallel-spin LLs the IQH
state (e.g., at $\nu=3$), remains strong. This behavior has been
attributed to easy-axis (for an opposite-spin crossing) and easy-plane
(for a same-spin crossing) ferromagnetism.
\cite{Muraki.PRL.2001,Vakili.PRL.2006,Jungwirth.PRL.1998}


We highlight three further observations. First, strong FQH states at
large $q/3$ fillings have been recently observed in very high quality
graphene samples.\cite{Dean.cond.mat.2010} These states qualitatively
resemble what we see in our two-subband system. It is tempting to
associate the valley degree of freedom in graphene with the subband
degree of freedom in our sample. But the LL structure in graphene is
of course different from GaAs so it is not obvious if this association
is valid. Second, data taken in the $N$ = 1 LL at very low
temperatures and in the highest quality, single-subband samples
exhibit FQH states at even-denominator fillings $\nu$ = 5/2 and 7/2.
\cite{Willett.PRL.1987,Pan.PRB.2008} In the traces shown in
Fig.~\ref{fig:waterfall}, we do not see any even-denominator states
when $N$ = 1, e.g., at $\nu$ = 7/2 in region B where $E_F$ is in the
S1$\uparrow$ level. However, in the same sample, at higher densities
($n > 3.4\times 10^{11}$ cm$^{-2}$) and at low temperatures ($T=$ 30
mK), we do indeed observe a FQH state at $\nu$ = 7/2 flanked by very
weak 10/3 and 11/3 states when $E_F$ lies in the S1$\uparrow$ level.
\cite{Shabani.PRL.2010}

Third, in the $N=0$ LL, high-quality samples show strong higher-order,
odd-denominator FQH states at composite Fermion filling factor
sequences such as 2/5, 3/7, 4/9, etc.\cite{CF_Jain} We do observe a
qualitatively similar behavior in our data when $E_F$ is in an $N$ = 0
LL. For example, in region A (Figs.~\ref{fig:waterfall} and
\ref{fig:colorful}(a)) we see weak but clear minima at $\nu$ = 17/5
next to the 10/3 minimum. Again, at higher densities and low
temperatures, such states become more developed.
\cite{Shabani.PRL.2010} In Fig. 1(b), for example, there are strong
minima at $\nu=$ 12/5 and 13/5, adjacent to the 7/3 and 8/3 minima,
and at 17/5 and 18/5, adjacent to the 10/3 and 11/3 minima. These
states, as well as the $q/3$ states, exhibit subtle evolutions even
when $E_F$ lies within a fixed $N=0$ LL, consistent with the presence
of composite Fermions which have spin and/or subband degrees of
freedom.\cite{Shabani.PRL.2010} A related question concerns the role
of charge distribution symmetry in the stability of the $q/3$
states. In other words, in a QW with fixed width, density and filling,
and with $E_F$ in a particular $N=0$ LL, how does the strength of
given a FQH state at a particular filling vary with charge
distribution symmetry. We do not have data to answer this question
quantitatively, but the data we present here clearly indicates that a
primary factor determining the strength of the $q/3$ FQH states is
whether or not $E_F$ lies in an $N=0$ LL.


\section{summary}

In conclusion, the position of $E_F$ is what determines the stability
of odd-denominator, $q/3$ FQH states at a given filling factor. When
$E_F$ lies in a ground-state ($N=0$) LL, the $q/3$ FQH states are
stable and strong, regardless of whether that LL belongs to the
symmetric or antisymmetric subband. This observation implies that the
wavefunction node in the out-of-plane direction is not detrimental to
the stability of these FQH states.  Also, the $q/3$ FQH states appear
to be stable very near the crossing of two LLs, especially if the LLs
have parallel spins.



\begin{acknowledgments}
  We acknowledge support through the NSF (DMR-0904117 and MRSEC
  DMR-0819860) for sample fabrication and characterization, and the
  DOE BES (DE-FG0200-ER45841) for measurements. We thank J. K. Jain
  and Z. Papic for illuminating discussions.
\end{acknowledgments}

\bibliography{paper_v1}

\end{document}